 \documentclass[noshowpacs,amsmath,twocolumn,
aps,prb]
{revtex4-1}

\usepackage{setspace}
\usepackage{amsmath}
\usepackage{graphicx}
\usepackage[nearskip,margin = 0pt]{subfig}
\usepackage{subfig}

\usepackage{verbatim}
\usepackage{amsfonts}
\usepackage{amssymb}
\usepackage{epstopdf} 
\usepackage{xcolor}
\DeclareGraphicsExtensions{.pdf,.eps,.png,.jpg,.mps} 
\usepackage{hyperref}
\hypersetup{
    colorlinks=true,
    linkcolor=blue,
    citecolor=blue,
    filecolor=blue,      
    urlcolor=blue,
}

\usepackage[utf8]{inputenc} 
\usepackage{caption}

\begin{document}

\title{Enabling high spectral efficiency coherent superchannel transmission\\ with soliton microcombs}


\author{Mikael Mazur$^{1\dagger}$, Myoung-Gyun Suh$^{2\dagger}$, Attila Fülöp$^{1}$, Jochen Schröder$^{1}$, Victor Torres-Company$^{1}$, Magnus Karlsson$^{1}$, Kerry J. Vahala$^{2}$ and Peter A. Andrekson$^{1}$\\
$^{1}$ Photonics Laboratory, Fibre Optic Communication Research Centre (FORCE), Department of Microtechnology and Nanoscience, Chalmers University  of Technology, Gothenburg  SE-412 96, Sweden\\
$^{2}$ T. J. Watson Laboratory of Applied Physics, California Institute of Technology, Pasadena, CA 91125, USA\\
$^\dagger$ These authors contributed equally to this work\\
 }

\maketitle

{\bf Optical communication systems have come through five orders of magnitude improvement in data rate over the last three decades\cite{Winzer2018}. 
The increased demand in data traffic and the limited optoelectronic component bandwidths\cite{Laperle2014} have led to state-of-the-art systems employing hundreds of separate lasers in each transmitter. Given the limited optical amplifier bandwidths, focus is now shifting to maximize the \emph{spectral efficiency}, SE \cite{Liu2014,Winzer2018}.
However, the frequency jitter from neighbouring lasers results in uncertainties of the exact channel wavelength, requiring large guardbands to avoid catastrophic channel overlap~\cite{Liu2018}.
Optical frequency combs with optimal line spacings (typically around 10-50\,GHz) can overcome these limitations and maximize the SE\cite{Lundberg2018}. Recent developments in microresonator-based soliton frequency combs (hereafter \emph{microcombs})\cite{Herr2013,Yi2015,Brasch2015,Wang2016,Joshi2016,Kippenberg2018} promise a compact, power efficient multi-wavelength and phase-locked light source for optical communications\cite{marin2017microresonator}. Here we demonstrate a microcomb-based communication link achieving state-of-the-art spectral efficiency that has previously only been possible with bulk-optics systems. 
Compared to previous microcomb works in optical communications\cite{marin2017microresonator,Fueloep2017}, our microcomb features a narrow line spacing of 22.1\,GHz. In addition, it provides a four order-of-magnitude more stable line spacing compared to free-running lasers~\cite{Liu2018}. The optical signal-to-noise ratio (OSNR) is sufficient for information encoding using state-of-the-art high-order modulation formats. This enables us to demonstrate transmission of a 12\,Tb/s superchannel over distances ranging from a single 82\,km span with an SE exceeding 10\,bits/s/Hz, to 2000\,km with an SE higher than 6\,bits/s/Hz. These results demonstrate that microcombs can attain the SE that will spearhead future optical networks. Combined with further advances in hybrid integration~\cite{Atabaki2018}, high-SE microcomb-based transmitters could enable novel transmission schemes with lower energy consumption~\cite{Lundberg2018} while continuing the decades of exponential growth in optical communications.}

\begin{figure}[b!]
\centering
\includegraphics[width=\linewidth]{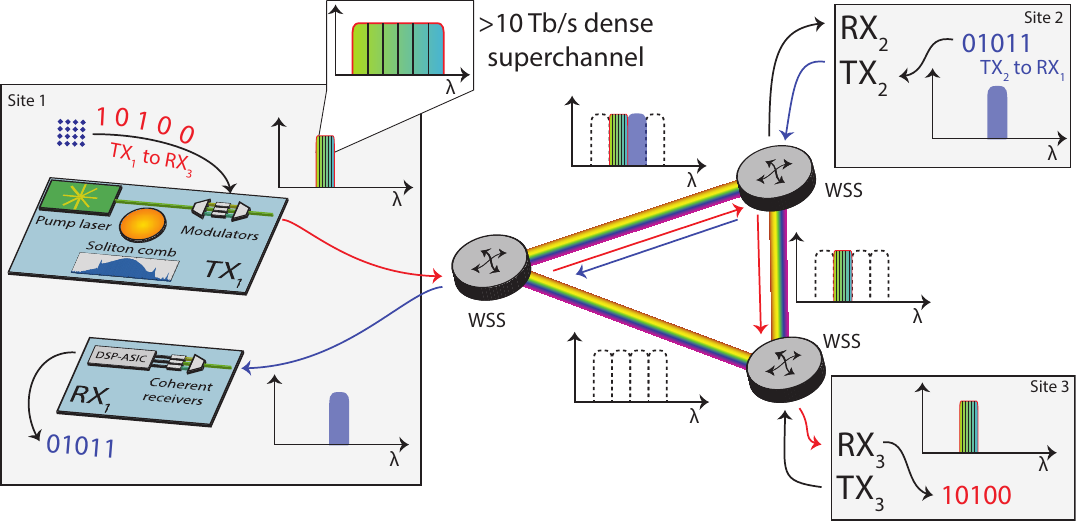}
\captionsetup{singlelinecheck=no, justification = RaggedRight}
\caption{{\bf Concept of microcomb-based superchannels in future networks.} In an all-optically routed network, superchannels can be routed arbitrarily across the entire system. They enable multi-Tb/s throughput with high spectral efficiency by avoiding excising guardbands between channels which are needed to separate individual channels using wavelength selective switches (WSS). The sketch shows an example of such a scenario using a chip-scale microcomb-based superchannel transmitter (Tx). The superchannel is transmitted from site 1 to a receiver (Rx) in site 3 (red) while a second superchannel goes from site 2 to site 1 (blue). High spectral line density microcombs allow the superchannels to be generated on a chip while maintaining state-of-the-art data spectral effiencies, thereby maximizing the throughput of the entire network.}
\label{fig:concept}
\end{figure}

Early optical communication systems encoded information in a binary-like fashion by switching the laser on and off. The invention of all-optical amplification using erbium-doped fibre amplifiers (EDFAs) provided a paradigm shift in the 90s by enabling simultaneous amplification over about $10\,$THz bandwidth, the C+L-bands. Throughput could be massively increased using wavelength-division multiplexing transmission (WDM). Today, high-speed digital-to-analogue and analogue-to-digital converters (DACs/ADCs)~\cite{Laperle2014} combined with digital signal processing (DSP) have enabled the adaptation of wireless coherent technologies in optical systems~\cite{Savory2008}. Classical on-off keying systems are being replaced with coherent solutions using advanced modulation formats~\cite{Savory2008,Winzer2018}. These formats exploit both amplitude and phase together with polarization multiplexing (PM) to transmit independent information. Moreover, DSP has replaced optical compensation of fibre dispersion~\cite{Savory2008}. 

\begin{figure*}[hbtp]
\centering
\includegraphics[width=\linewidth]{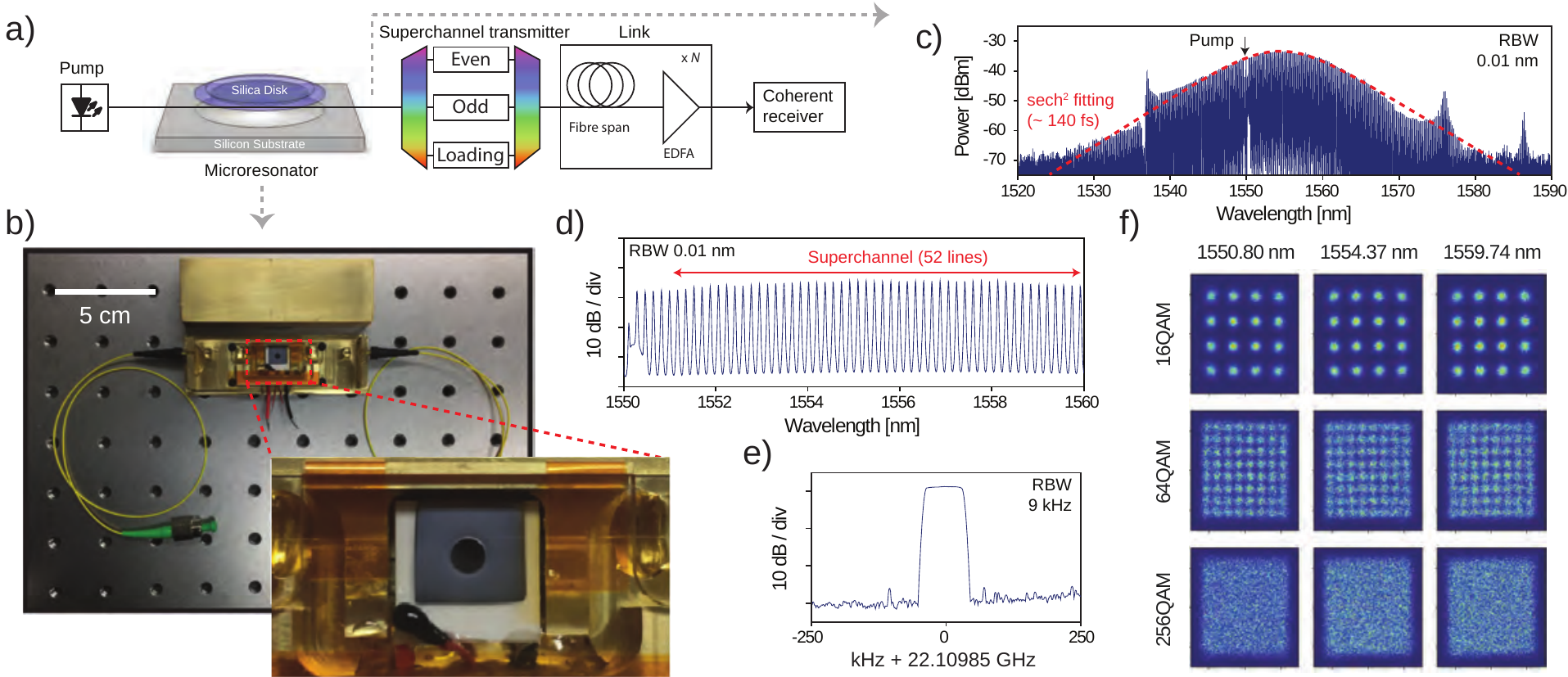}
\captionsetup{singlelinecheck=no, justification = RaggedRight}
\caption{{\bf Coherent superchannel transceiver using soliton microcombs.} a. Schematic experimental setup used to perform the superchannel transmission. A single wavelength continuous-wave (CW) laser was amplified via an erbium-doped fibre amplifier (EDFA) and coupled into a microresonator to generate a soliton microcomb. The high-Q silica microresonator consists of a whispering-gallery-mode silica wedge disk resonator (blue) on a silicon substrate (gray). After suppressing the pump laser line using a fibre Bragg grating, the soliton microcomb was amplified, flattened, and demultiplexed. A test-loading-band approach with 5 test channels was used to emulate individual modulators for each channel. After modulation, the channels are combined again via a multiplexer. The optical signal was launched into a single span fibre loop or a recirculating loop. Finally, the signal was demultiplexed and detected with a coherent receiver. b. Picture of the packaged microresonator module. Inset: Zoomed-in view of 22.1 GHz soliton microcomb device. c. Optical spectrum of the soliton microcomb. The squared hyperbolic secant envelope (dashed red curve) gives a soliton pulse width of 140 fs. d. Zoom-in of the optical spectrum showing the superchannel (52 comb lines) used in the transmission experiment. e. Electrical spectrum (measured for $>$3 hours in maxhold mode) showing  the long-term stability of the 22.1 GHz repetition rate. f. Constellation diagrams of 16, 64, 256QAM at three different wavelengths after 82\,km of superchannel transmission (X-polarization). 
}
\label{fig:setup}
\end{figure*}

The throughput demand of single transceivers is rapidly approaching 10\,Tb/s~\cite{Winzer2017}. At the same time the electrical bandwidth, determining the maximum symbol rate and the performance of the transceiver electronics, is increasing at a much lower pace. Previously, sufficient excess fibre bandwidth, limited by the EDFA gain, has been available to use standard WDM transmission with poor spectral efficiency to scale beyond the limitations of a single transceiver. As this bandwidth is rapidly filling up and optical networks are changing from simple point-to-point to optically routed networks supporting flexible channel bandwidths, the growth from WDM has stagnated. One promising path to enable continued growth is to introduce superchannels, as conceptually illustrated in Fig.~\ref{fig:concept}. A superchannel is a high-throughput channel formed by optical multiplexing of multiple classical channels. It can therefore reach orders of magnitude higher bandwidths than the individual channels. Superchannels are crucial to maximize the efficiency of future optical networks supporting flexible channel widths and use all-optical routing~\cite{Winzer2018}. Achieving similar throughput using classical WDM channels would lead to poor performance in flexible optically routed networks. This follows from the limitations of optical routers which require guardbands exceeding 10\,GHz between 50\,GHz-spaced channels to avoid substantial performance penalty from channel-selective filtering using optical wavelength selective switches (WSSes)~\cite{Fabrega2016}. As the exact guard-band required depends on the number of nodes, the penalty increase with network  flexibility. However, since a superchannel is viewed as a single broadband channel, it is always routed as a unit and there is no need to have any routing guardbands between each channel. The minimum guard-band between the channels forming the superchannel is therefore mainly dictated by the uncertainty in absolute laser frequency~\cite{Lundberg2018}.

\begin{figure*}[htbp]
\centering
\includegraphics[width=\linewidth]{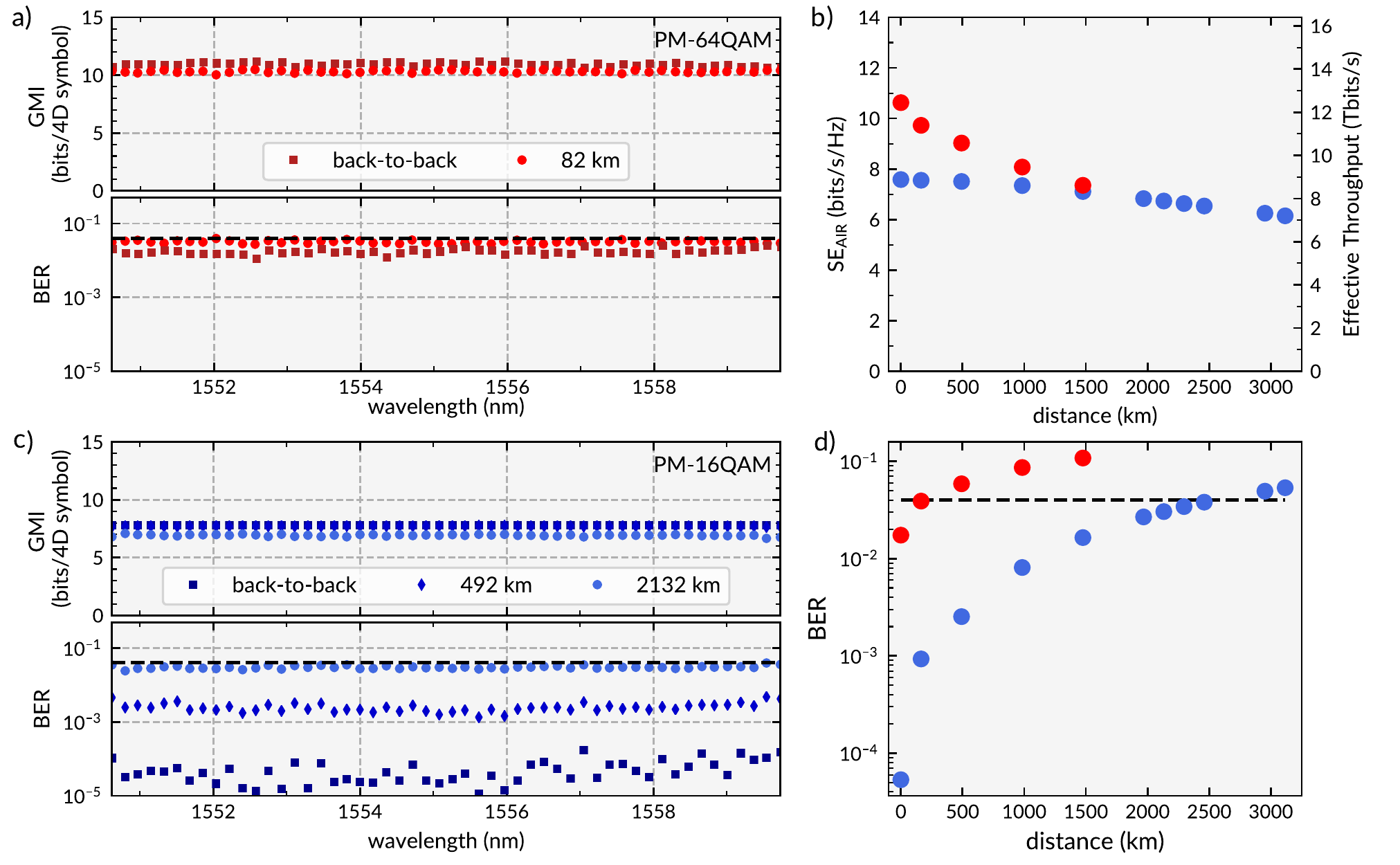}
\captionsetup{singlelinecheck=no, justification = RaggedRight}
\caption{{\bf Superchannel transmission results.} The superchannel performance is measured using both bit error rate (BER) and generalized mutual information (GMI). All measurements were done at optimal launch power corresponding to about 13.5\,dBm. The dashed line indicates the target BER threshold (see methods) a. GMI and BER for all channels for back-to-back and 82\,km transmission. All 52 channels were modulated with 21.5\,Gbaud PM-64QAM with a resulting spectral efficiency of 10.1\,bits/s/Hz. The corresponding superchannel throughput was 11.7\,Tb/s. b. Resulting BER and GMI for all wavelength channels in back-to-back, after 490\,km and after 2150\,km of transmission using a recirculating loop. Each channel was modulated with PM-16QAM resulting in a spectral efficiency of 6.7\,bits/s/Hz or equivalent 8.0\,Tb/s superchannel throughput after 2150\,km c. Spectral efficiency and throughput vs. transmission distance ($N\times 164$\,km). Red dots denote PM-64QAM modulation and blue PM-16QAM. d. Average superchannel BER as a function of distance for PM-16QAM (blue) and PM-64QAM (red) transmission. }
\label{fig:txresult}
\end{figure*}

Using independent lasers, the minimum required guard-band is a few GHz\cite{Liu2018,Rahn2018}. This limitation can lead to loss in efficiency of about 10\% for systems targeting symbol rates around 30\,Gbaud~\cite{Rahn2018}. The guard-band penalty is thus the strongest factor limiting system data rates. High symbol rates help to mitigate this problem, but at the expense of increasing the penalty associated with DAC and ADC imperfections~\cite{Laperle2014}. To maximize the spectral efficiency, the gaps must be minimized which can only be done using mutually frequency locked optical carriers. One attractive way to achieve this is to use an optical frequency comb that provides multiple phase-locked carriers~\cite{Hall2006}. Frequency combs are the basis of modern frequency metrology, time keeping and spectroscopy \cite{Diddams2010} and as a WDM source in optical communications they can replace many independent lasers thereby reducing the system energy consumption~\cite{marin2017microresonator,Fueloep2018}. Beyond maximizing spectral efficiency, frequency locking between the lines of a frequency comb can also enable novel DSP concepts to improve performance~\cite{Temprana2015} and/or reduce energy consumption~\cite{Lundberg2018}. However, traditional frequency combs are bulky with high power consumption. 

The recent demonstration of soliton microcombs represents a major turning point for the ubiquitous comb applications (e.g. frequency synthesis\cite{Spencer2018}, spectroscopy\cite{Suh2016,Dutt2018}, laser-ranging \cite{Trocha2018,Suh2018a} and optical clocks~\cite{Newman2018}) because of their compact form factor and low operating power~\cite{Stern2018,Volet2018}. In particular, soliton formation ensures highly stable mode locking and reproducible spectral envelopes\cite{Leo2010,Herr2013}. Soliton microcombs have also been used for data transmission, reaching a combined throughput of 50\,Tb/s over a single-hop 75\,km link\cite{marin2017microresonator}. The transmitter in that experiment used two interleaved 100\,GHz spaced combs to reach a resulting spectral efficiency of 5.2\,bits/s/Hz. When using a single comb, the spectral efficiency was decreased to 2.8\,bits/s/Hz. To improve the spectral efficiency, lower frequency spacing (typically 10-50 GHz) is required. The larger mode volume of such devices, however, increases the operational power of the soliton microcombs at these low repetition rates. This increased operational power can be offset by leveraging the inverse quadratic dependence of parametric oscillation threshold power on optical Q factor\cite{Kippenberg2004} as has been demonstrated using crystalline and silica-based microcombs\cite{Herr2013,Yi2015,Suh2018,Yang2018}.

In this experiment, a high-Q whispering-gallery-mode silica resonator \cite{Lee2012}(Fig. 2a) was used to generate a soliton microcomb with 22.1 GHz repetition rate. The device was packaged in a compact module (Fig. 2b) for better thermal stability and portability (see Methods). Figure 2c shows the optical spectrum of the generated soliton. While the comb was pumped at 1549.9\,nm, we took advantage of the Raman self-frequency shift\cite{Yi2015,Yi2016,Karpov2016} to displace the spectral center of the soliton envelope to near 1555\,nm. This enabled selection of 52 carriers around 1555\,nm without having to sacrifice any carriers around the pump notch filter, crucial to maximizing the system efficiency. The zoomed-in optical spectrum in Fig. 2d shows the equidistantly-spaced comb lines over 10\,nm span measured after amplification, which were used for the transmission experiment. The electrical spectrum (Fig. 2e) shows that the soliton repetition rate of 22.1\,GHz is stable within 100\,kHz range over several hours. A temperature feedback of the microcomb module allowed continuous operation of the soliton microcomb throughout the entire transmission experiment over 100 hours (see Methods).

To demonstrate the feasibility of high spectral efficiency microcomb-based superchannels we conducted two transmission experiments. Following laboratory constraints, we evaluated the concept using the common test-loading-band approach to ensure realistic carrier OSNR and minimize any penalty from not having independent modulators for each channel~\cite{puttnam20152}. The experimental setup is shown in Fig.~\ref{fig:setup}a (see methods and extended Fig. 1 for more details). The superchannel was formed using 52 lines from the microcomb.

We modulated each carrier at 21.5\,Gbaud using PM-$M$-ary quadrature amplitude modulation (PM-$M$QAM) with $M\in[16,64,256]$. Both the modulation order and the symbol rate  were optimized (see methods) to maximize the system throughput. In a first demonstration we transmitted the superchannel over a single-hop 82\,km link, as shown in Fig.~\ref{fig:txresult}a. The combined superchannel reached a spectral efficiency of 10.1\,bits/s/Hz using PM-64QAM while assuming soft-decision forward error correction (FEC) (see Methods). This could be slightly increased to 10.4\,bits/s/Hz using PM-256QAM at the expense of increased coding complexity. The corresponding throughput was 11.7\,Tb/s (12.0\,Tb/s) for PM-64QAM (PM-256QAM) with a total channel bandwidth of about 1.1\,THz. Constellation diagrams for three different wavelengths after 82\,km of transmission are shown in Fig.~\ref{fig:setup}f. These results are directly comparable with state-of-the-art bulk comb sources~\cite{Mazur2018b,puttnam20152}. 

To further investigate the capability of the microcomb-based superchannel we used a recirculating loop to significantly increase the transmission distance. The results for both PM-16QAM and PM-64QAM are shown in Fig.~\ref{fig:txresult}b. PM-64QAM results in the highest spectral efficiency (throughput) but comes at the expense of requiring more powerful FEC, similar to  PM-256QAM in the single-span measurement. Measured performance for each line using PM-16QAM after 2150\,km of transmission is shown in Fig.~\ref{fig:txresult}c. All lines are below reference thresholds~\cite{Schuh2017} for the assumed FEC, verifying that reaching beyond 2000\,km of transmission is feasible by changing from PM-64QAM to PM-16QAM. At this distance, the spectral efficiency was 6.7\,bits/s/Hz, corresponding to a throughput of 8.0\,Tb/s. Average superchannel BER for both PM-16QAM and PM-64QAM is shown in Fig.~\ref{fig:txresult}d. 

The combination of these two results highlights the capability of our microcomb to cope with the various requirements for future transceivers.  While distance covered here spans the range from 80 to 2000\,km, even longer distances can be reached by expanding the modulation alphabet with formats such as PM-4QAM and PM-8QAM or probabilistic shaping on constellation formats with higher cardinality~\cite{JaureguiRuiz2018}. 


In summary, we have demonstrated the highest spectral efficiency superchannel transmission using a microcomb light source. Combination of the optimal line spacing and the unique stability of soliton microcombs enabled spectral efficiencies exceeding 10\,bits/s/Hz in a single-span 82\,km link and transmission over distances ranging more than 2500\,km with spectral efficiency exceeding 6\,bits/s/Hz. Rapidly progressing microcomb research has resulted in microcombs operating at lower operational power\cite{Liu2018a,Suh2018,Volet2018}, and at much reduced footprint through integration with diode lasers\cite{Stern2018,Raja2018} and simplification of microcomb triggering systems\cite{Pavlov2018}. These advances will greatly reduce the power consumption and form factor of microcombs, and will eventually allow microcomb-based superchannels to spear-head future highly-flexible all-optical networks requiring multi-Tb/s throughput. 
%
\vspace{3 mm}

\noindent\textbf{Methods}

\medskip
\begin{footnotesize}
\noindent\textbf{Microresonator and device module.}
The high-Q silica microresonator featured a 3 mm diameter corresponding to a 22.1 GHz comb line spacing. Device fabrication process and typical performance characteristics including quality factors and parametric oscillation threshold for this design are reported elsewhere \cite{Lee2012,Suh2018}. The device was packaged in a compact module (30 mm x 94 mm x 15 mm) that includes fibre pigtails and a thermoelectric cooling (TEC) element. The device module was fabricated at the California Institute of Technology (Pasadena, CA, USA) and shipped to Chalmers University of Technology (Göteborg, Sweden) for the optical transmission experiment.\\

\noindent\textbf{Comb initialization and stabilization.}
A 1550\,nm fibre laser (Ethernal SlowLight Laser manufactured by Orbits Lightwave, Inc.) was amplified by an EDFA to approximately 150\,mW and coupled into the packaged device module. The pump laser was rapidly (at a speed of $\sim$ 300 GHz/s) tuned into the soliton existence detuning range for soliton triggering. After the transmitted pump power was filtered using a fibre bragg grating, the soliton power was tapped via a 99/1 fibre splitter. The one percent port of soliton power was photodetected and used as an error signal for servo control to maintain the soliton state.

To improve the long-term stability of the system, the device temperature was controlled via a TEC element installed inside the module. With the temperature feedback, the soliton microcomb was stable throughout the entire transmission experiment over 100 hours. To measure the absolute stability of the microcomb offset frequency, the central comb line was beat with a commercial self-referenced fibre comb (Menlo FC1500-250-ULN) and observed every minute over a time span of multiple days showing standard deviation of $<$ 5 MHz.\\

\noindent\textbf{Transmission experiment.}
In order to encode information on the carriers and form the complete superchannel, the comb output was first amplified before a wavelength selective switch (WSS) (Finisar Waveshaper 4000S) separated the lines into three paths. To enable accurate evaluation without artificial OSNR degradation we used the common test-loading-band approach~\cite{puttnam20152} to cope with the laboratory limitation of only having access to 3 modulators. The test-band consisted of 5 channels and was generated from two WSS outputs. The WSS outputs were individually amplified and modulated using the even-odd channel principle to ensure uncorrelated cross-talk for the channels under test. Out of the 5 channels in the test-band we only measured the central 3 to ensure proper OSNR of the neighbouring channels. A sketch of the experimental setup can be found in the extended Fig. 1. 

The IQ-modulators used were driven by 2 independent DACs each (Keysight M8195A), operating at about 60\,GS/s to modulate the 21.5\,Gbaud signal shaped with a root-raised cosine filter with 1\% roll-off. Symbol rate optimization was used to maximize the resulting spectral efficiency by balancing overhead and linear cross-talk from non-ideal electrical components in the transmitter. We evaluated every PM-$M$QAM format with $M\in[8,16,32,64,128,256,512,1024]$ from which we selected the square formats with $M\in[16,64,256]$ to maximize the SE for both short and long transmission distances (Extended Fig. 2.). The transmitter had an implementation penalty of about 1\,dB for PM-64QAM and 0.5\,dB for PM-16QAM (Extended Fig. 3.). The test-band outputs were then combined, amplified and polarization multiplexing was emulated using the split-delay-combine method with a delay of $\approx$ 250 symbols. The remaining channels were modulated in a single modulator and optical dispersive decorrelation was used to emulate the case of independent data on each channel. A similar PM emulation stage was used to emulate PM signals. After post-modulation amplification, a second WSS (Finisar Waveshaper 1000S) was used to create a notch at the test-band position and ensure a flat loading band. The test and loading channels were then combined to form the final superchannel. 

The link consisted of either a single span of 82\,km standard single-mode fibre (SSMF) or two consecutive spans placed inside a recirculating loop. The loss of each span was about 16.5\,dB. The loss was compensated using EDFAs with about 5.5\,dB noise figure. The recirculating loop was controlled using two acusto-optic modulators and light was coupled in and out using a 3\,dB coupler. A WSS and a 10\,nm bandpass filter were placed inside to ensure a flat spectrum and to filter out-of-band amplifier noise every roundtrip. In addition, a polarization scrambler was used to randomly vary the input state of polarization to the fist span. We optimized the launch power for both single span and loop transmission with resulting optima found to be 13.5\,dBm for both single span using PM-64QAM and loop transmission using PM-16QAM with 2000\,km target distance. At the output, the signal was amplified before selecting the channel under test with a 0.25\,nm bandpass filter. A standard coherent receiver with 23\,GHz analogue bandwidth was used to measure the signal after down-conversion with a free-running external cavity laser with maximum 100\,kHz specified linewidth. The local oscillator frequency was tuned to be within $\pm100$\,MHz from the corresponding comb line. The resulting electrical signals were sampled with a 50\,GS/s real-time oscilloscope and processed offline. The DSP was pilot-aided~\cite{Mazur2018b} (available open-source~\cite{Schroeder2018}),with 1.5\% and 2.6\% overhead for single span and loop measurements respectively. No data-aided pre-convergence or adaptation using the payload symbols was performed. For the long-haul measurements using PM-16QAM, a blind phase algorithm was used in addition to the pilot-based algorithm.\\

\noindent\textbf{Spectral efficiency and performance metrics.}
Measuring the received bit error ratio (BER) is traditionally used as a straight forward metric for system performance. By setting a target BER level, often around $10^{-3} - 10^{-2}$, and assuming a suitable forward error-correcting code (FEC) with a corresponding required data overhead, one can quantify a net data rate. The direct one-to-one relationship between pre-FEC BER and post-FEC data rate breaks down, however, when using modern soft-decision FEC codes (often at pre-FEC BERs above $10^{-2}$)\cite{Alvarado2016}. Therefore, it has become customary to also calculate the generalized mutual information (GMI) in addition to the pre-FEC BER for systems employing high-order modulation formats. In addition to being compatible with soft-decision FECs, the GMI allows us to avoid setting an arbitrary target level that all channels must conform to, resembling the adaptive coding principles used in modern systems~\cite{Ghazisaeidi2015}. The net data rate for each channel is instead calculated assuming an optimized coding overhead for that particular channel. We should note that GMI is a lower bound on the classical capacity in an additive white Gaussian noise channel, given by the mutual information (MI), constraining the available codes to bit-interleaved coded-modulation schemes. The achievable information rate (AIR)-values presented here are therefore lower than the maximum AIR assuming MI-based coding.\cite{Alvarado2016}

To evaluate the performance of our system we measured and calculated both the BER and the GMI. In this work we use a BER threshold of $4\cdot10^{-2}$, previously used in for PM-64QAM transmission assuming 20\% FEC data overhead~\cite{Schuh2017}. Our spectral efficiency estimates are however based on the GMI metric, in line with recent high spectral efficiency experiments. Both the BER and GMI are calculated using $>10$ million bits per wavelength channel. The spectral efficiency is then calculated from the GMI by deducting additional overheads from the guardbands (2.8\%) and DSP pilots, resulting in a GMI-based achievable information rate (AIR). 





\vspace{3 mm}

\noindent \textbf{Funding Information.}
Chalmers funding from the Swedish research council (VR) and the European research council (grant 771410). Caltech funding from the Air Force Office of Scientific Research (FA9550-18-1-0353) and the Kavli Nanoscience Institute.

\noindent\textbf{Competing Interests.}
The authors declare no competing interests.
\vspace{1 mm}

\noindent\textbf{Author Information.} Correspondence and requests for materials should be addressed to P. A. A. (email: peter.andrekson@chalmers.se) or K. J. V. (email: vahala@caltech.edu).

\bibliographystyle{naturemag}
\bibliography{main}

\begin{thebibliography}{10}
\expandafter\ifx\csname url\endcsname\relax
  \def\url#1{\texttt{#1}}\fi
\expandafter\ifx\csname urlprefix\endcsname\relax\def\urlprefix{URL }\fi
\providecommand{\bibinfo}[2]{#2}
\providecommand{\eprint}[2][]{\url{#2}}

\bibitem{Winzer2018}
\bibinfo{author}{Winzer, P.~J.}, \bibinfo{author}{Neilson, D.~T.} \&
  \bibinfo{author}{Chraplyvy, A.~R.}
\newblock \bibinfo{title}{Fiber-optic transmission and networking: the previous
  20 and the next 20 years [invited]}.
\newblock \emph{\bibinfo{journal}{Optics Express}}
  \textbf{\bibinfo{volume}{26}}, \bibinfo{pages}{24190--24239}
  (\bibinfo{year}{2018}).

\bibitem{Laperle2014}
\bibinfo{author}{Laperle, C.} \& \bibinfo{author}{Osullivan, M.}
\newblock \bibinfo{title}{{Advances in high-speed DACs, ADCs, and DSP for
  optical coherent transceivers}}.
\newblock \emph{\bibinfo{journal}{Journal of Lightwave Technology}}
  \textbf{\bibinfo{volume}{32}}, \bibinfo{pages}{629--643}
  (\bibinfo{year}{2014}).

\bibitem{Liu2014}
\bibinfo{author}{Liu, X.}, \bibinfo{author}{Chandrasekhar, S.} \&
  \bibinfo{author}{Winzer, P.~J.}
\newblock \bibinfo{title}{Digital signal processing techniques enabling
  multi-{Tb/s} superchannel transmission: An overview of recent advances in
  {DSP}-enabled superchannels}.
\newblock \emph{\bibinfo{journal}{{IEEE} Signal Processing Magazine}}
  \textbf{\bibinfo{volume}{31}}, \bibinfo{pages}{16--24}
  (\bibinfo{year}{2014}).

\bibitem{Liu2018}
\bibinfo{author}{Liu, G.} \emph{et~al.}
\newblock \bibinfo{title}{Demonstration of a carrier frequency offset estimator
  for 16-/32-{QAM} coherent receivers: a hardware perspective}.
\newblock \emph{\bibinfo{journal}{Optics Express}}
  \textbf{\bibinfo{volume}{26}}, \bibinfo{pages}{4853--4862}
  (\bibinfo{year}{2018}).

\bibitem{Lundberg2018}
\bibinfo{author}{Lundberg, L.} \emph{et~al.}
\newblock \bibinfo{title}{Frequency comb-based {WDM} transmission systems
  enabling joint signal processing}.
\newblock \emph{\bibinfo{journal}{Applied Sciences}}
  \textbf{\bibinfo{volume}{8}}, \bibinfo{pages}{718} (\bibinfo{year}{2018}).

\bibitem{Herr2013}
\bibinfo{author}{Herr, T.} \emph{et~al.}
\newblock \bibinfo{title}{Temporal solitons in optical microresonators}.
\newblock \emph{\bibinfo{journal}{Nature Photonics}}
  \textbf{\bibinfo{volume}{8}}, \bibinfo{pages}{145--152}
  (\bibinfo{year}{2013}).

\bibitem{Yi2015}
\bibinfo{author}{Yi, X.}, \bibinfo{author}{Yang, Q.-F.}, \bibinfo{author}{Yang,
  K.~Y.}, \bibinfo{author}{Suh, M.-G.} \& \bibinfo{author}{Vahala, K.}
\newblock \bibinfo{title}{Soliton frequency comb at microwave rates in a high-q
  silica microresonator}.
\newblock \emph{\bibinfo{journal}{Optica}} \textbf{\bibinfo{volume}{2}},
  \bibinfo{pages}{1078--1085} (\bibinfo{year}{2015}).

\bibitem{Brasch2015}
\bibinfo{author}{Brasch, V.} \emph{et~al.}
\newblock \bibinfo{title}{Photonic chip-based optical frequency comb using
  soliton cherenkov radiation}.
\newblock \emph{\bibinfo{journal}{Science}} \textbf{\bibinfo{volume}{351}},
  \bibinfo{pages}{357--360} (\bibinfo{year}{2015}).

\bibitem{Wang2016}
\bibinfo{author}{Wang, P.-H.} \emph{et~al.}
\newblock \bibinfo{title}{Intracavity characterization of micro-comb generation
  in the single-soliton regime}.
\newblock \emph{\bibinfo{journal}{Optics Express}}
  \textbf{\bibinfo{volume}{24}}, \bibinfo{pages}{10890--10897}
  (\bibinfo{year}{2016}).

\bibitem{Joshi2016}
\bibinfo{author}{Joshi, C.} \emph{et~al.}
\newblock \bibinfo{title}{Thermally controlled comb generation and soliton
  modelocking in microresonators}.
\newblock \emph{\bibinfo{journal}{Optics Letters}}
  \textbf{\bibinfo{volume}{41}}, \bibinfo{pages}{2565--2568}
  (\bibinfo{year}{2016}).

\bibitem{Kippenberg2018}
\bibinfo{author}{Kippenberg, T.~J.}, \bibinfo{author}{Gaeta, A.~L.},
  \bibinfo{author}{Lipson, M.} \& \bibinfo{author}{Gorodetsky, M.~L.}
\newblock \bibinfo{title}{Dissipative kerr solitons in optical
  microresonators}.
\newblock \emph{\bibinfo{journal}{Science}} \textbf{\bibinfo{volume}{361}},
  \bibinfo{pages}{eaan8083} (\bibinfo{year}{2018}).

\bibitem{marin2017microresonator}
\bibinfo{author}{Marin-Palomo, P.} \emph{et~al.}
\newblock \bibinfo{title}{{Microresonator-based solitons for massively parallel
  coherent optical communications}}.
\newblock \emph{\bibinfo{journal}{Nature}} \textbf{\bibinfo{volume}{546}},
  \bibinfo{pages}{274--279} (\bibinfo{year}{2017}).

\bibitem{Fueloep2017}
\bibinfo{author}{Fülöp, A.} \emph{et~al.}
\newblock \bibinfo{title}{Long-haul coherent communications using
  microresonator-based frequency combs}.
\newblock \emph{\bibinfo{journal}{Optics Express}}
  \textbf{\bibinfo{volume}{25}}, \bibinfo{pages}{26678--26688}
  (\bibinfo{year}{2017}).

\bibitem{Atabaki2018}
\bibinfo{author}{Atabaki, A.~H.} \emph{et~al.}
\newblock \bibinfo{title}{Integrating photonics with silicon nanoelectronics
  for the next generation of systems on a chip}.
\newblock \emph{\bibinfo{journal}{Nature}} \textbf{\bibinfo{volume}{556}},
  \bibinfo{pages}{349--354} (\bibinfo{year}{2018}).

\bibitem{Savory2008}
\bibinfo{author}{Savory, S.~J.}
\newblock \bibinfo{title}{Digital filters for coherent optical receivers}.
\newblock \emph{\bibinfo{journal}{Optics Express}}
  \textbf{\bibinfo{volume}{16}}, \bibinfo{pages}{804--817}
  (\bibinfo{year}{2008}).

\bibitem{Winzer2017}
\bibinfo{author}{Winzer, P.~J.} \& \bibinfo{author}{Neilson, D.~T.}
\newblock \bibinfo{title}{From scaling disparities to integrated parallelism: A
  decathlon for a decade}.
\newblock \emph{\bibinfo{journal}{Journal of Lightwave Technology}}
  \textbf{\bibinfo{volume}{35}}, \bibinfo{pages}{1099--1115}
  (\bibinfo{year}{2017}).

\bibitem{Fabrega2016}
\bibinfo{author}{Fabrega, J.~M.} \emph{et~al.}
\newblock \bibinfo{title}{On the filter narrowing issues in elastic optical
  networks}.
\newblock \emph{\bibinfo{journal}{Journal of Optical Communications and
  Networking}} \textbf{\bibinfo{volume}{8}}, \bibinfo{pages}{A23--A33}
  (\bibinfo{year}{2016}).

\bibitem{Rahn2018}
\bibinfo{author}{Rahn, J.} \emph{et~al.}
\newblock \bibinfo{title}{{DSP}-enabled frequency locking for near-nyquist
  spectral efficiency superchannels utilizing integrated photonics}.
\newblock In \emph{\bibinfo{booktitle}{Proceedings of Optical Fiber
  Communication Conference}} (\bibinfo{year}{2018}).
\newblock \bibinfo{note}{Paper W1B.3}.

\bibitem{Hall2006}
\bibinfo{author}{Hall, J.~L.}
\newblock \bibinfo{title}{Nobel lecture: Defining and measuring optical
  frequencies}.
\newblock \emph{\bibinfo{journal}{Reviews of Modern Physics}}
  \textbf{\bibinfo{volume}{78}}, \bibinfo{pages}{1279--1295}
  (\bibinfo{year}{2006}).

\bibitem{Diddams2010}
\bibinfo{author}{Diddams, S.~A.}
\newblock \bibinfo{title}{The evolving optical frequency comb [invited]}.
\newblock \emph{\bibinfo{journal}{Journal of the Optical Society of America B}}
  \textbf{\bibinfo{volume}{27}}, \bibinfo{pages}{B51} (\bibinfo{year}{2010}).

\bibitem{Fueloep2018}
\bibinfo{author}{Fülöp, A.} \emph{et~al.}
\newblock \bibinfo{title}{High-order coherent communications using mode-locked
  dark-pulse kerr combs from microresonators}.
\newblock \emph{\bibinfo{journal}{Nature Communications}}
  \textbf{\bibinfo{volume}{9}}, \bibinfo{pages}{1598} (\bibinfo{year}{2018}).

\bibitem{Temprana2015}
\bibinfo{author}{Temprana, E.} \emph{et~al.}
\newblock \bibinfo{title}{Overcoming kerr-induced capacity limit in optical
  fiber transmission}.
\newblock \emph{\bibinfo{journal}{Science}} \textbf{\bibinfo{volume}{348}},
  \bibinfo{pages}{1445--1448} (\bibinfo{year}{2015}).

\bibitem{Spencer2018}
\bibinfo{author}{Spencer, D.~T.} \emph{et~al.}
\newblock \bibinfo{title}{An optical-frequency synthesizer using integrated
  photonics}.
\newblock \emph{\bibinfo{journal}{Nature}} \textbf{\bibinfo{volume}{557}},
  \bibinfo{pages}{81--85} (\bibinfo{year}{2018}).

\bibitem{Suh2016}
\bibinfo{author}{Suh, M.-G.}, \bibinfo{author}{Yang, Q.-F.},
  \bibinfo{author}{Yang, K.~Y.}, \bibinfo{author}{Yi, X.} \&
  \bibinfo{author}{Vahala, K.~J.}
\newblock \bibinfo{title}{Microresonator soliton dual-comb spectroscopy}.
\newblock \emph{\bibinfo{journal}{Science}} \textbf{\bibinfo{volume}{354}},
  \bibinfo{pages}{600--603} (\bibinfo{year}{2016}).

\bibitem{Dutt2018}
\bibinfo{author}{Dutt, A.} \emph{et~al.}
\newblock \bibinfo{title}{On-chip dual-comb source for spectroscopy}.
\newblock \emph{\bibinfo{journal}{Science Advances}}
  \textbf{\bibinfo{volume}{4}}, \bibinfo{pages}{e1701858}
  (\bibinfo{year}{2018}).

\bibitem{Trocha2018}
\bibinfo{author}{Trocha, P.} \emph{et~al.}
\newblock \bibinfo{title}{Ultrafast optical ranging using microresonator
  soliton frequency combs}.
\newblock \emph{\bibinfo{journal}{Science}} \textbf{\bibinfo{volume}{359}},
  \bibinfo{pages}{887--891} (\bibinfo{year}{2018}).

\bibitem{Suh2018a}
\bibinfo{author}{Suh, M.-G.} \& \bibinfo{author}{Vahala, K.~J.}
\newblock \bibinfo{title}{Soliton microcomb range measurement}.
\newblock \emph{\bibinfo{journal}{Science}} \textbf{\bibinfo{volume}{359}},
  \bibinfo{pages}{884--887} (\bibinfo{year}{2018}).

\bibitem{Newman2018}
\bibinfo{author}{Newman, Z.~L.} \emph{et~al.}
\newblock \bibinfo{title}{Photonic integration of an optical atomic clock.}
  \bibinfo{note}{Preprint at arXiv}, \eprint{1811.00616v1}.

\bibitem{Stern2018}
\bibinfo{author}{Stern, B.}, \bibinfo{author}{Ji, X.},
  \bibinfo{author}{Okawachi, Y.}, \bibinfo{author}{Gaeta, A.~L.} \&
  \bibinfo{author}{Lipson, M.}
\newblock \bibinfo{title}{Battery-operated integrated frequency comb
  generator}.
\newblock \emph{\bibinfo{journal}{Nature}} \textbf{\bibinfo{volume}{562}},
  \bibinfo{pages}{401--405} (\bibinfo{year}{2018}).

\bibitem{Volet2018}
\bibinfo{author}{Volet, N.} \emph{et~al.}
\newblock \bibinfo{title}{Micro-resonator soliton generated directly with a
  diode laser}.
\newblock \emph{\bibinfo{journal}{Laser {\&} Photonics Reviews}}
  \textbf{\bibinfo{volume}{12}}, \bibinfo{pages}{1700307}
  (\bibinfo{year}{2018}).

\bibitem{Leo2010}
\bibinfo{author}{Leo, F.} \emph{et~al.}
\newblock \bibinfo{title}{Temporal cavity solitons in one-dimensional kerr
  media as bits in an all-optical buffer}.
\newblock \emph{\bibinfo{journal}{Nature Photonics}}
  \textbf{\bibinfo{volume}{4}}, \bibinfo{pages}{471--476}
  (\bibinfo{year}{2010}).

\bibitem{Kippenberg2004}
\bibinfo{author}{Kippenberg, T.~J.}, \bibinfo{author}{Spillane, S.~M.} \&
  \bibinfo{author}{Vahala, K.~J.}
\newblock \bibinfo{title}{Kerr-nonlinearity optical parametric oscillation in
  an ultrahigh-{QToroid} microcavity}.
\newblock \emph{\bibinfo{journal}{Physical Review Letters}}
  \textbf{\bibinfo{volume}{93}}, \bibinfo{pages}{083904}
  (\bibinfo{year}{2004}).

\bibitem{Suh2018}
\bibinfo{author}{Suh, M.-G.} \& \bibinfo{author}{Vahala, K.}
\newblock \bibinfo{title}{Gigahertz-repetition-rate soliton microcombs}.
\newblock \emph{\bibinfo{journal}{Optica}} \textbf{\bibinfo{volume}{5}},
  \bibinfo{pages}{65--66} (\bibinfo{year}{2018}).

\bibitem{Yang2018}
\bibinfo{author}{Yang, K.~Y.} \emph{et~al.}
\newblock \bibinfo{title}{Bridging ultrahigh-q devices and photonic circuits}.
\newblock \emph{\bibinfo{journal}{Nature Photonics}}
  \textbf{\bibinfo{volume}{12}}, \bibinfo{pages}{297--302}
  (\bibinfo{year}{2018}).

\bibitem{Lee2012}
\bibinfo{author}{Lee, H.} \emph{et~al.}
\newblock \bibinfo{title}{Chemically etched ultrahigh-{Q} wedge-resonator on a
  silicon chip}.
\newblock \emph{\bibinfo{journal}{Nature Photonics}}
  \textbf{\bibinfo{volume}{6}}, \bibinfo{pages}{369--373}
  (\bibinfo{year}{2012}).

\bibitem{Yi2016}
\bibinfo{author}{Yi, X.}, \bibinfo{author}{Yang, Q.-F.}, \bibinfo{author}{Yang,
  K.~Y.} \& \bibinfo{author}{Vahala, K.}
\newblock \bibinfo{title}{Theory and measurement of the soliton self-frequency
  shift and efficiency in optical microcavities}.
\newblock \emph{\bibinfo{journal}{Optics Letters}}
  \textbf{\bibinfo{volume}{41}}, \bibinfo{pages}{3419--3422}
  (\bibinfo{year}{2016}).

\bibitem{Karpov2016}
\bibinfo{author}{Karpov, M.} \emph{et~al.}
\newblock \bibinfo{title}{Raman self-frequency shift of dissipative kerr
  solitons in an optical microresonator}.
\newblock \emph{\bibinfo{journal}{Physical Review Letters}}
  \textbf{\bibinfo{volume}{116}}, \bibinfo{pages}{103902}
  (\bibinfo{year}{2016}).

\bibitem{puttnam20152}
\bibinfo{author}{Puttnam, B.} \emph{et~al.}
\newblock \bibinfo{title}{{2.15 Pb/s transmission using a 22 core homogeneous
  single-mode multi-core fiber and wideband optical comb}}.
\newblock In \emph{\bibinfo{booktitle}{Proceedings of European Conference on
  Optical Communication}} (\bibinfo{year}{2015}).
\newblock \bibinfo{note}{Paper. PDP3.1}.

\bibitem{Mazur2018b}
\bibinfo{author}{Mazur, M.}, \bibinfo{author}{Lorences-Riesgo, A.},
  \bibinfo{author}{Schroder, J.}, \bibinfo{author}{Andrekson, P.~A.} \&
  \bibinfo{author}{Karlsson, M.}
\newblock \bibinfo{title}{10 {Tb/s} {PM}-{64QAM} self-homodyne comb-based
  superchannel transmission with 4{\%} shared pilot tone overhead}.
\newblock \emph{\bibinfo{journal}{Journal of Lightwave Technology}}
  \textbf{\bibinfo{volume}{36}}, \bibinfo{pages}{3176--3184}
  (\bibinfo{year}{2018}).

\bibitem{Schuh2017}
\bibinfo{author}{Schuh, K.} \emph{et~al.}
\newblock \bibinfo{title}{Single carrier 1.2 tbit/s transmission over 300 km
  with {PM}-64 {QAM} at 100 {GBaud}}.
\newblock In \emph{\bibinfo{booktitle}{Proceedings of Optical Fiber
  Communication Conference}} (\bibinfo{publisher}{{OSA}},
  \bibinfo{year}{2017}).
\newblock \bibinfo{note}{Paper Th5B.5}.

\bibitem{JaureguiRuiz2018}
\bibinfo{author}{de~Jauregui~Ruiz, I.~F.} \emph{et~al.}
\newblock \bibinfo{title}{25.4-{Tb/s} transmission over transpacific distances
  using truncated probabilistically shaped {PDM}-{64QAM}}.
\newblock \emph{\bibinfo{journal}{Journal of Lightwave Technology}}
  \textbf{\bibinfo{volume}{36}}, \bibinfo{pages}{1354--1361}
  (\bibinfo{year}{2018}).

\bibitem{Liu2018a}
\bibinfo{author}{Liu, J.} \emph{et~al.}
\newblock \bibinfo{title}{Ultralow-power chip-based soliton microcombs for
  photonic integration}.
\newblock \emph{\bibinfo{journal}{Optica}} \textbf{\bibinfo{volume}{5}},
  \bibinfo{pages}{1347--1353} (\bibinfo{year}{2018}).

\bibitem{Raja2018}
\bibinfo{author}{Raja, A.~S.} \emph{et~al.}
\newblock \bibinfo{title}{Electrically pumped photonic integrated soliton
  microcomb.} \bibinfo{note}{Preprint at arXiv}, \eprint{1810.03909}.

\bibitem{Pavlov2018}
\bibinfo{author}{Pavlov, N.~G.} \emph{et~al.}
\newblock \bibinfo{title}{Narrow-linewidth lasing and soliton kerr microcombs
  with ordinary laser diodes}.
\newblock \emph{\bibinfo{journal}{Nature Photonics}}
  \textbf{\bibinfo{volume}{12}}, \bibinfo{pages}{694--698}
  (\bibinfo{year}{2018}).

\bibitem{Schroeder2018}
\bibinfo{author}{Schroeder, J.} \& \bibinfo{author}{Mazur, M.}
\newblock \bibinfo{title}{Chalmersphotonicslab/qampy: V0.1}
  (\bibinfo{year}{2018}).
\newblock \bibinfo{note}{Https://github.com/ChalmersPhotonicsLab/QAMpy}.

\bibitem{Alvarado2016}
\bibinfo{author}{Alvarado, A.}, \bibinfo{author}{Agrell, E.},
  \bibinfo{author}{Lavery, D.}, \bibinfo{author}{Maher, R.} \&
  \bibinfo{author}{Bayvel, P.}
\newblock \bibinfo{title}{Replacing the soft-decision {FEC} limit paradigm in
  the design of optical communication systems}.
\newblock \emph{\bibinfo{journal}{Journal of Lightwave Technology}}
  \textbf{\bibinfo{volume}{34}}, \bibinfo{pages}{707--721}
  (\bibinfo{year}{2016}).

\bibitem{Ghazisaeidi2015}
\bibinfo{author}{Ghazisaeidi, A.} \emph{et~al.}
\newblock \bibinfo{title}{Transoceanic transmission systems using adaptive
  multirate {FECs}}.
\newblock \emph{\bibinfo{journal}{Journal of Lightwave Technology}}
  \textbf{\bibinfo{volume}{33}}, \bibinfo{pages}{1479--1487}
  (\bibinfo{year}{2015}).

\end{thebibliography}

\end{footnotesize}


\begin{figure*}[hbtp]
\centering
\includegraphics[width=\textwidth]{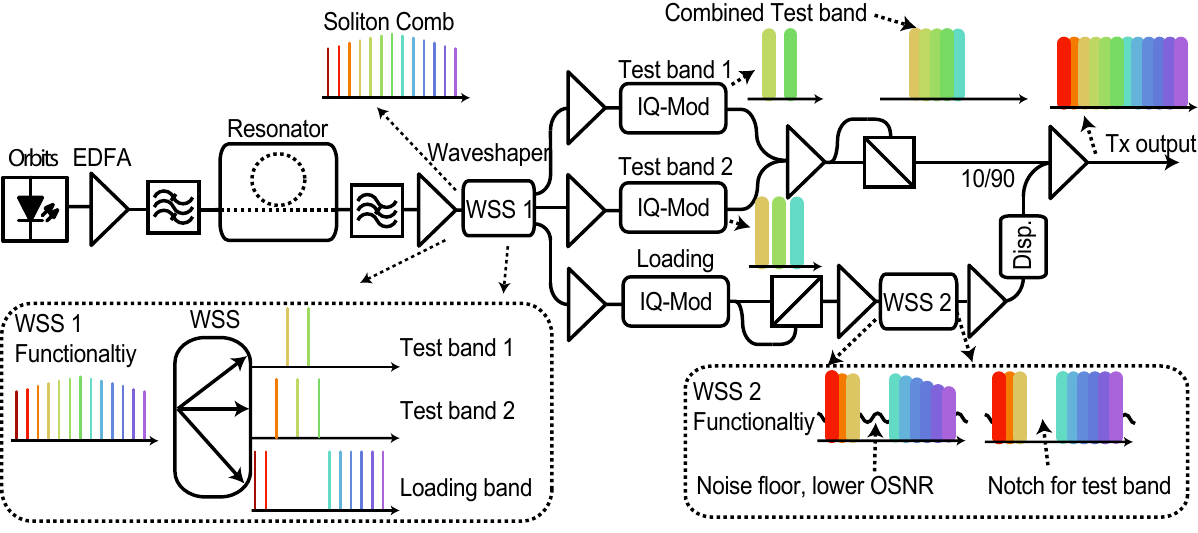}
\captionsetup{singlelinecheck=no, justification = RaggedRight}
\caption*{{\bf Extended FIG. 1.} Experimental transmitter setup (described in Methods) to generate the high spectral efficiency superchannel. A test-loading-band approach was used to avoid excess OSNR degradation from laboratory constraints originating from only having 3 IQ-modulators. The test-band consisted of 5 channels modulated using the even-odd principle. The loading band was modulated using a third modulator and the individual channels were optically decorrelated to avoid enhanced nonlinear penalties.}
\end{figure*}

\begin{figure*}[hbtp]
\centering
\includegraphics[width=\textwidth]{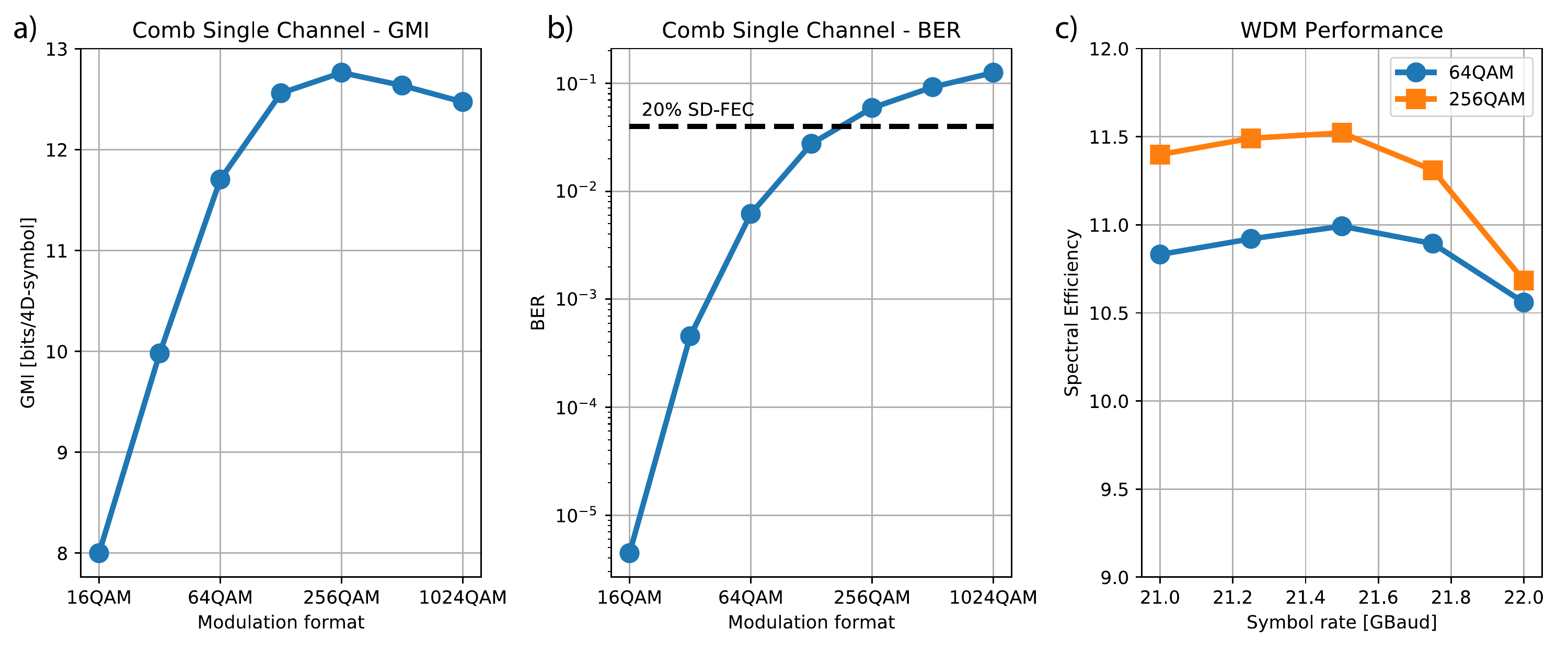}
\captionsetup{singlelinecheck=no, justification = RaggedRight}
\caption*{{\bf Extended FIG. 2.} Back-to-back optimization of modulation formats and symbol rate to maximime the transmitter performance. a GMI as a function of modulation format for PM-$M$QAM with $M\in[16,32,64,128,256,512,1024]$. The highest GMI is observed for PM-256QAM, reaching about 12.7\,bits/4D-symbol. The performance GMI for PM-64QAM was about 11.7\,bits/4D-symbol. b BER as a function of modulation order. With respect to a, we observe that while PM-256QAM results in the highest information rate, it also requires stronger SD-FEC overheads than the about 20\% normally considered. We therefore choose PM-64QAM as the main modulation format for short-reach (although we also measured using PM-256QAM as a reference). c Optimization of symbol rate to balance overhead from guard-bands and penalty from electrical cross-talk. A symbol rate of 21.5\,GBaud on the 22.1\,GHz grid provided by the comb resulted in the highest spectral efficiency and was used throughput the transmission experiments. The corresponding overhead for the 600\,MHz guard-band was 2.7\%. 
 }
\end{figure*}

\begin{figure*}[hbtp]
\centering
\includegraphics[width=\textwidth]{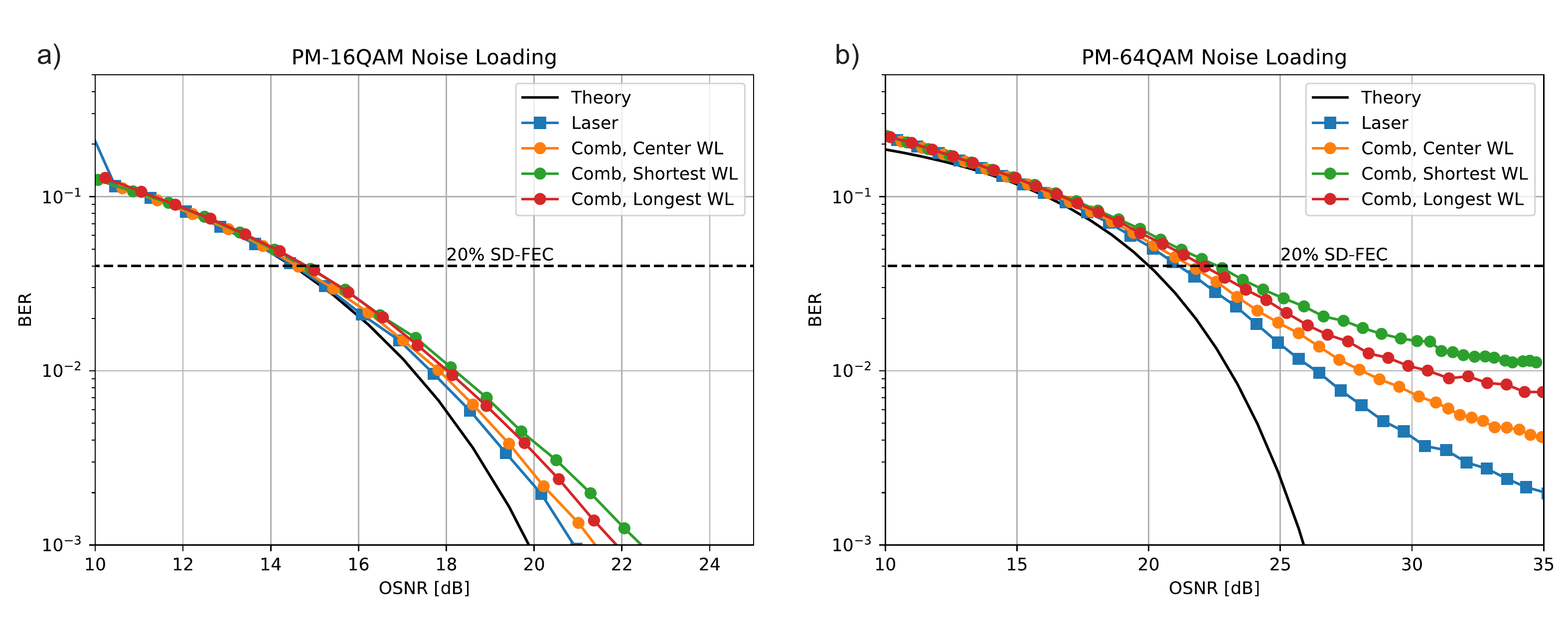}
\captionsetup{singlelinecheck=no, justification = RaggedRight}
\caption*{{\bf Extended FIG. 3.} Noise loading measurements to quantify implementation penalty for a, 21.5\,Gbaud PM-16QAM and b, PM-64QAM. The implementation penalty is quantified with respect to theory at the target threshold of $4\cdot 10^{-2}$ (see Methods). For 21.5\,Gbaud PM-64QAM, we observe an implementation penalty of about 1\,dB with respect to theory when only using the comb seed laser. In addition, the penalty from using the frequency comb ranges from $<0.5$\,dB to about 1.5\,dB. The highest penalty is observed for the shortest wavelength, being about 0.5\,dB higher than the longest used wavelength. This was due to varying gain and noise figure of the EDFA used to amplify the comb output. The corresponding penalties was about 0.5\,dB for PM-16QAM when using only the comb seed laser. Adding the microcomb resulted in a maximum additional penalty of around 0.5\,dB.}
\end{figure*}

\end{document}